\documentclass[twocolumn,preprintnumbers,amsmath,amssymb,groupedaddress,superscriptaddress,prr]{revtex4-1}

\usepackage[utf8]{inputenc}
\usepackage{graphicx}
\usepackage{dcolumn}
\usepackage[colorlinks=true,citecolor=blue,urlcolor=blue]{hyperref}
\usepackage[dvipsnames]{xcolor}
\usepackage{subfigure}

\usepackage{amsmath,amssymb,cases} 
\usepackage{bm}
\newcommand{\eqnref}

\begin{document}
\title{Observation of Quantum Droplets in a Heteronuclear Bosonic Mixture}

\author{C. D'Errico}\email{derrico@lens.unifi.it}
\affiliation{Istituto Nazionale di Ottica, CNR-INO, 50019 Sesto Fiorentino, Italy}
\affiliation{\mbox{LENS and Dipartimento di Fisica e Astronomia, Universit\`{a} di Firenze, 50019 Sesto Fiorentino, Italy}}

\author{A. Burchianti}\email{burchianti@lens.unifi.it}
\affiliation{Istituto Nazionale di Ottica, CNR-INO, 50019 Sesto Fiorentino, Italy}
\affiliation{\mbox{LENS and Dipartimento di Fisica e Astronomia, Universit\`{a} di Firenze, 50019 Sesto Fiorentino, Italy}}

\author{M. Prevedelli}
\affiliation{Dipartimento di Fisica e Astronomia, Universit\`{a} di Bologna, 40127 Bologna, Italy}

\author{L. Salasnich}
\affiliation{Istituto Nazionale di Ottica, CNR-INO, 50019 Sesto Fiorentino, Italy}
\affiliation{\mbox{Dipartimento di Fisica e Astronomia 'Galileo Galilei' and CNISM, Universit\`{a} di Padova, 35131 Padova, Italy}}

\author{F. Ancilotto}
\affiliation{\mbox{Dipartimento di Fisica e Astronomia 'Galileo Galilei' and CNISM, Universit\`{a} di Padova, 35131 Padova, Italy}}
\affiliation{CNR-IOM Democritos, 265–34136 Trieste, Italy}

\author{M. Modugno}
\affiliation{\mbox{Depto. de Fis\'ica Te\'orica e Hist. de la Ciencia, Universidad del Pais Vasco UPV/EHU, 48080 Bilbao, Spain}}
\affiliation{IKERBASQUE, Basque Foundation for Science, 48013 Bilbao, Spain}

\author{F. Minardi}
\affiliation{Istituto Nazionale di Ottica, CNR-INO, 50019 Sesto Fiorentino, Italy}
\affiliation{\mbox{LENS and Dipartimento di Fisica e Astronomia, Universit\`{a} di Firenze, 50019 Sesto Fiorentino, Italy}}
\affiliation{Dipartimento di Fisica e Astronomia, Universit\`{a} di Bologna, 40127 Bologna, Italy}

\author{C. Fort}
\affiliation{Istituto Nazionale di Ottica, CNR-INO, 50019 Sesto Fiorentino, Italy}
\affiliation{\mbox{LENS and Dipartimento di Fisica e Astronomia, Universit\`{a} di Firenze, 50019 Sesto Fiorentino, Italy}}


\begin{abstract}
We report on the formation of heteronuclear quantum droplets in an attractive bosonic mixture of $^{41}$K and $^{87}$Rb. We observe long-lived self-bound states, both in free space and in an optical waveguide. In the latter case, the dynamics under the effect of a species-dependent force confirms their bound nature. By tuning the interactions from the weakly to the strongly attractive regime, we study the transition from expanding to localized states, in both geometries. We compare the experimental results with beyond mean-field theory and we find a good agreement in the full range of explored interactions. Our findings open up the production of long-lived droplets with important implications for further research.
\end{abstract}

   \maketitle
\section{Introduction}
Interactions are ubiquitous in nature and understanding their effects is a primary challenge in physics, especially in many-body quantum systems. Generally, mean-field (MF) theories well reproduce a plethora of phenomena related to interparticle interactions. However, this approach fails when quantum fluctuations are non-negligible. The first-order correction to the MF energy, the so called Lee-Huang-Yang (LHY) term, was first calculated in 1957 \cite{LeeHuangYang} to describe bosons with hard-core repulsion. In ultracold gases the LHY term is normally negligible and only recently it has been experimentally investigated \cite{Altmeyer2007,Shin2008,Papp2008,Navon2010,Navon2011}.
Notwithstanding, there are situations where the LHY and MF contributions can be of the same order. For instance, in attractive bosonic mixtures the repulsive LHY term may stabilize the system against collapse, leading to the formation of self-bound droplets. Although very dilute, these states have a liquid-like behaviour, characterized by a core with uniform density in the large atom number limit \cite{Petrov}.

While originally predicted for Bose-Bose mixtures, quantum droplets have been first realized in dipolar gases, by exploiting the competition between contact repulsion and anisotropic dipole-dipole attraction \cite{kadau2016,FerrierBarbut2016,Schmitt2016,Ferlaino2016,FerrierBarbutJPB2016,Wenzel2017}. 
Recently, tuning contact interactions through a Feshbach resonance, quantum droplets have been also observed in a spin mixture of $^{39}$K, both in the presence of an external potential \cite{Cheiney2018,Cabrera2018} and in free space \cite{Semeghini2018}.
By studying collisions between droplets it has been possible to observe the crossover between compressible and incompressible regimes \cite{Ferioli2019}. Free space $^{39}$K droplets have a short lifetime (only few ms), limited by three body losses \cite{Semeghini2018,Ferioli2019}, motivating the quest for stable long-lived droplets in different atomic mixtures. Longer lifetimes could indeed allow the investigation of many peculiar features of these states, like the characterization of the incompressible regime and the observation of self-evaporation \cite{Petrov}.

In this work, we study the formation of quantum droplets in a heteronuclear Bose-Bose mixture of $^{41}$K and $^{87}$Rb where the two species experience different trapping potentials. We exploit Feshbach resonances for tuning the interspecies interaction. In particular, we explore the transition from the weakly to the strongly attractive regime, beyond the MF threshold for collapse, where self-bound quantum droplets are predicted to exist.
We observe that in the absence of any external trapping potential, the strongly attractive mixture remains localized on a timescale of several tens of milliseconds, consistently with the formation of long-lived self-bound droplets. Furthermore, we study the size evolution and the center-of-mass dynamics of the mixture in a horizontal optical waveguide, in the presence of a species-dependent magnetic force. Also in this geometry, where both the LHY correction and the radial confinement provide a mechanism to prevent the collapse \cite{Shabat72}, we observe the formation of a localized bound state, whose center-of-mass follows a trajectory intermediate between those of two weakly attractive $^{41}$K and $^{87}$Rb clouds.
By studying the dynamics, we extract the number ratio of $^{41}$K and $^{87}$Rb atoms forming the bound state, and we find it consistent with the value predicted by the theory \cite{Petrov}. Our experimental findings are well reproduced by numerical simulations performed by using two coupled generalized Gross-Pitaevskii (GP) equations  \cite{Gross,Pitaevskii}, including the LHY correction for heteronuclear mixtures \cite{Petrov,Ancilotto2018}.
\section{Experiment}
\begin{figure}[t]
\includegraphics[width = 0.46 \textwidth]{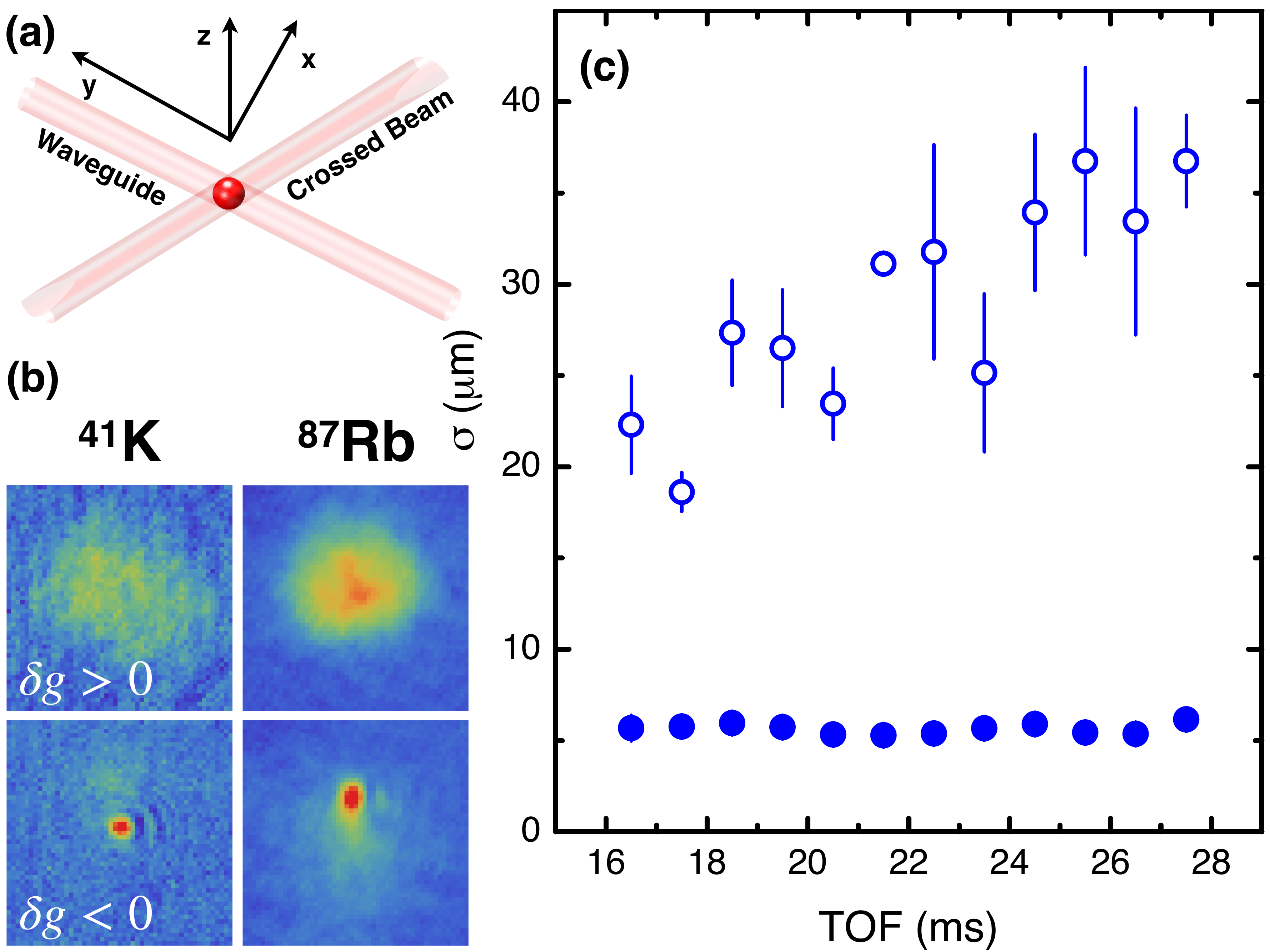}
\hspace{5mm} 
\caption{\label{fig:TOF} (a) Schematics of the geometry of the crossed trap. (b) Absorption images in the $xz$ plane ($190\;\mu$m x $190\;\mu$m) after $\mathrm{TOF}= 24.5$~ms ($27$ ms) for $^{41}$K ($^{87}$Rb) condensate in the weakly-attractive (top row) and in the strongly-attractive (bottom row) regimes. $^{41}$K atoms are imaged at the target Feshbach field, while $^{87}$Rb atoms are imaged at $B=0$~G after an additional TOF of $2.5$~ms. (c) Average size $\sigma = \sqrt{\sigma_x \sigma_z}$ of $^{41}$K cloud as a function of the TOF in free space for two values of the interspecies scattering length: $a_\mathrm{KRb} = (-17.5\pm1.4)  a_0$ (open circles), and $a_\mathrm{KRb} = (-84.5\pm1.6) a_0$ (closed circles). The error bars correspond to the standard deviation of, typically, five independent measurements.}
\end{figure}

We produce a binary condensate of $^{41}$K and $^{87}$Rb both prepared in the $\left|F=2,m_{F}=2\right\rangle$ state in a crossed optical dipole trap \cite{Burchianti2018}. 
To tune the interspecies interaction, two microwave pulses transfer both species in the $\left|F=1,m_{F}=1\right\rangle$ state, and a vertical homogeneous magnetic field is increased to $\simeq 71$~G, lying between two Feshbach resonances \cite{Thalhammer2008}. Due to their different masses, $^{41}$K and $^{87}$Rb atoms have a differential vertical gravitational sag and their equilibrium positions are separated by $\sim 15~\mu$m in the optical trap. 
We compensate for this offset by compressing the dipole trap and by adding a vertical magnetic field gradient $b_z=-16$~G/cm \cite{Hansen2013,Lous2017}, exploiting the different magnetic moments of the two species ($\mu_\mathrm{K} = -0.83 \;\mu_B$ and $\mu_\mathrm{Rb}= -0.52\;\mu_{B}$, at $70$~G), as shown in Appendix \ref{AppendixA}.
At this stage, we normally have a binary condensate with $N_\mathrm{K} \simeq 1 - 2 \times 10^4$ and $N_\mathrm{Rb} \simeq 4 - 6 \times 10^4$ atoms, confined in an approximately harmonic potential with different frequencies for the two atomic species: ($\nu_{x},\nu_y, \nu_z) \simeq (130, 90, 180)$~Hz for $^{41}$K and 
$(100, 70, 130)$~Hz for $^{87}$Rb. 
The homonuclear and heteronuclear scattering lengths are respectively $a_\mathrm{K}=65 a_0$ \cite{DErrico2007}, $a_\mathrm{Rb}=100.4 a_0$ \cite{Marte2002}, and $a_\mathrm{KRb} \simeq 0 a_0$ \cite{Simoni2008,Thalhammer2009}.  To enter the attractive regime, we then change the magnetic field to the desired value $B_F$ in $30$~ms, thus tuning $a_\mathrm{KRb}$, as detailed in Appendix \ref{AppendixA}.
The mixture is completely characterized in terms of the intraspecies $g_{\mathrm{K}}=4\pi \hbar^2 a_{\mathrm{K}}/m_{\mathrm{K}}$, $g_{\mathrm{Rb}}=4\pi \hbar^2 a_{\mathrm{Rb}}/m_{\mathrm{Rb}}$ and interspecies $g_{\mathrm{KRb}}=2\pi \hbar^2 a_{\mathrm{KRb}}/m_r$ coupling constants,  with $m_{\mathrm{K}}$ and $m_{\mathrm{Rb}}$ the atomic masses and $m_r$  the reduced mass. The onset of the MF collapse regime corresponds to $\delta g = g_\mathrm{KRb}+\sqrt{g_\mathrm{K} g_\mathrm{Rb}}=0$ ($a_\mathrm{KRb}=-75.4a_0)$ \cite{Riboli2002}. We explore the crossover from the weak ($a_\mathrm{KRb}<0$, $\delta g >0$) to the strong ($\delta g <0$) attractive regime.
For the measurements in free space, the trap beams and the magnetic field gradient $b_z$ are simultaneously switched off and the atoms are imaged in time-of-flight (TOF). For the expansion in the horizontal waveguide (along the $\hat{y}$ direction) one of the two beams of the dipole trap (labeled as crossed beam in Fig.~\ref{fig:TOF}a) 
is slowly turned off by linearly reducing its intensity in $10$~ms, and the atoms are imaged {\it in-situ} after a given evolution time. In the waveguide the atoms experience a radial trapping potential with average frequencies $\nu_{r} \simeq 160$~Hz ($120$~Hz)  and an axial anti-trapping potential, produced by the magnetic field gradient, with $\nu_{y} \simeq -1.8$~Hz ($-0.6$~Hz) for $^{41}$K ($^{87}$Rb).
\section{Results}
\subsection{Dynamics in free space}
We first consider the dynamics in free space.
In Fig.~\ref{fig:TOF}b we show typical absorption images of the two condensates in the $xz$ plane and $\mathrm{TOF}= 24.5$~ms for $\delta g>0$ (top) and $\delta g<0$ (bottom). 
In the former case both $^{41}$K and $^{87}$Rb samples freely expand, while in the latter case we observe that a fraction of both species bounds and forms a small and dense component. The different shape of $^{41}$K and $^{87}$Rb clouds for $\delta g<0$ is a consequence of our imaging procedure.
To detect both atomic species, in fact, we first take an absorption image of $^{41}$K sample at the magnetic field $B_F$, then we switch off $B_F$ and after further $2.5$~ms of TOF we take the image of $^{87}$Rb atoms. 
The actual size of the bound state can be measured only in the first image, after which the state is dissociated and $^{87}$Rb cloud expands.
In Fig. \ref{fig:TOF}c we plot the average width (rms) $\sigma = \sqrt{\sigma_x \sigma_z}$ of $^{41}$K sample as a function of TOF for both interaction regimes. Whereas for $\delta g>0$ (open circles) $\sigma$ increases in time, for $\delta g<0$ (closed circles) it does not exceed our imaging resolution ($5\;\mu$m) up to $28$~ms of TOF, indicating the formation of a droplet state. 
In our mixture the droplet density is expected to be one order of magnitude smaller than in $^{39}$K at the same $\delta g$, leading to a substantial reduction of the three-body losses. This is consistent with a two orders of magnitude longer lifetime. Such difference is mainly due to the larger scattering lengths of $^{41}$K and $^{87}$Rb, since the density scales as $n \sim (\delta g/g)^2 a^{-3}$ for approximately equal scattering lengths, as discussed in Appendix \ref{AppendixB}.
In the experiment, for TOF $>28$~ms, we observe that the atomic clouds start to expand. We attribute this to the variation of $a_{\mathrm{KRb}}$, since for larger TOF the atoms fall off the region where the Feshbach field is spatially homogeneous. Nevertheless, this time is longer than the maximum observation time of $^{39}$K droplets in free space \cite{Semeghini2018}.
In order to increase the observation time, one could exploit a magnetic gradient $b_z$ to levitate the droplet, with an average magnetic moment per particle:
\begin{equation}
\mu_\mathrm{KRb} = \frac{N^d_\mathrm{K} \mu_\mathrm{K}+N^d_\mathrm{Rb} \mu_\mathrm{Rb}}{N^d_\mathrm{K}+N^d_\mathrm{Rb}},
\end{equation}
where $N^d_\mathrm{K}$ and $N^d_\mathrm{Rb}$ are the number of $^{41}$K and $^{87}$Rb atoms forming the bound state.
However, we observed that, due to the species-dependent magnetic force, the droplet breaks up for $b_z\gtrsim 5$~G/cm.
Thus, we apply a lower magnetic field gradient $b_z=1.1$~G/cm, that, even if too small to compensate for gravity, reduces the variation of the Feshbach field during TOF \footnote{After $25$~ms of TOF we estimate a variation of the Feshbach field of $0.17$~G, resulting in a $a_{\mathrm{KRb}}$ variation of $2.9 a_0$.}. In addition, it introduces a species-dependent acceleration sufficient to spatially discriminate the bound and the unbound atoms. 
In the images at $\delta g<0$ (Fig. \ref{fig:TOF}b bottom), we note indeed the presence of a halo surrounding the localized clouds of both $^{41}$K and $^{87}$Rb. Due to the finite temperature, a fraction of atoms remains unbound. Moreover, for an efficient sympathetic cooling of the mixture, the atom number of $^{87}$Rb is typically three or four times larger than that of $^{41}$K, leading to an excess of unbound $^{87}$Rb atoms. For $^{41}$K the unbound halo is upshifted  ($|\mu_\mathrm{K}|>|\mu_\mathrm{KRb}|$), whereas for $^{87}$Rb is downshifted ($|\mu_\mathrm{Rb}|<|\mu_\mathrm{KRb}|$).
By measuring the accelerations of the different components one could extract $\mu_{\mathrm{KRb}}$ and therefore the ratio of the atom numbers $N^d_\mathrm{K}/N^d_\mathrm{Rb}$ in the droplet. 
However, in free space, the acceleration is dominated by gravity (for the droplet the ratio between magnetic and gravitational forces is only $0.035$).
\subsection{Waveguide dynamics}
\begin{figure}[t]
\includegraphics[width = 0.46\textwidth]{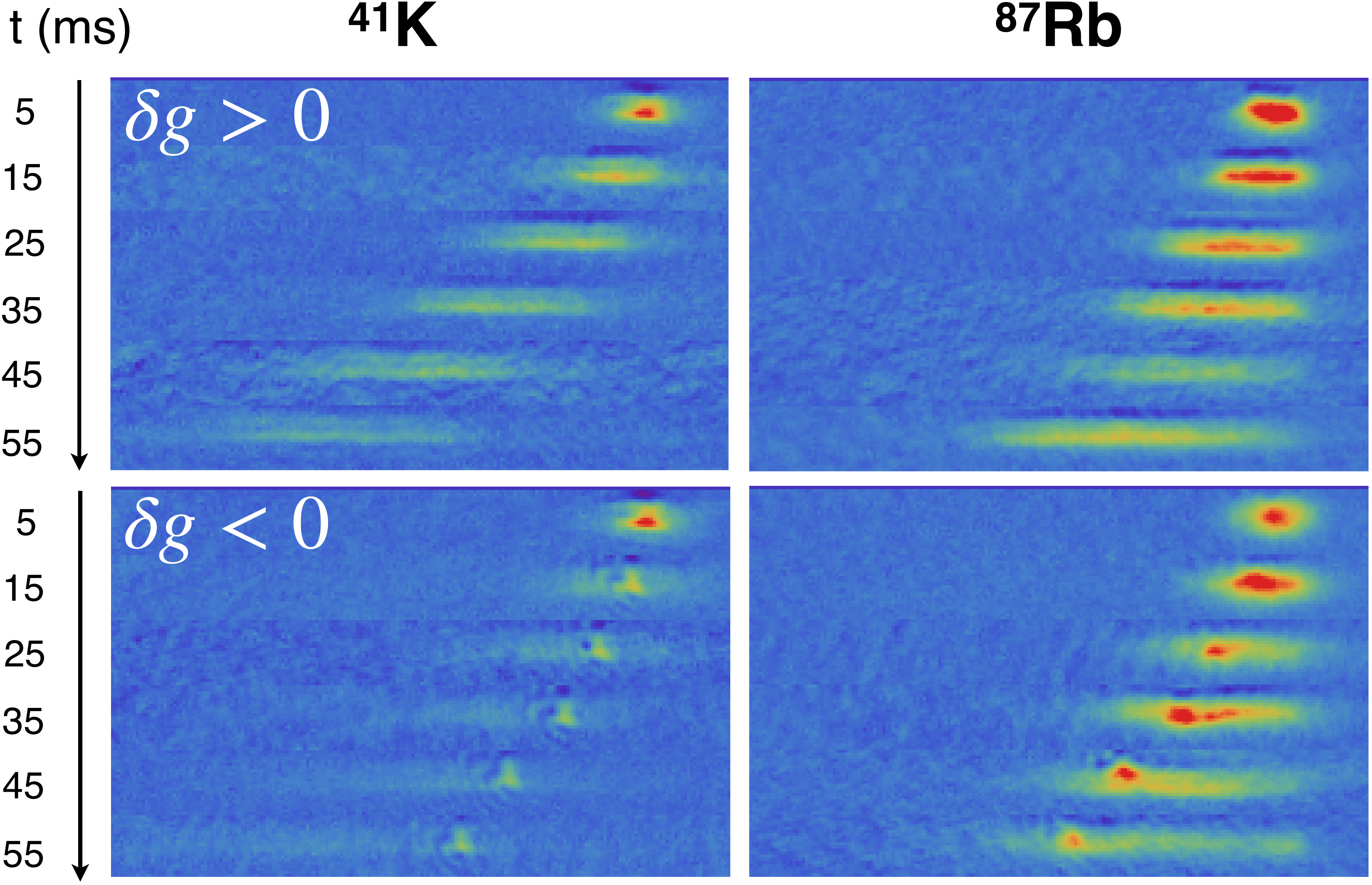}
\caption{\label{fig:ImmaginiEspansioneGuida} \textit{In-situ} absorption images in the $yz$ plane of the atoms expanding in a horizontal waveguide ($760\;\mu$m x $266\;\mu$m) at increasing (from top to bottom) evolution time $t$ in the weakly attractive regime $a_\mathrm{KRb} = (-11.4\pm1.4) a_0$ (top row) and in the strongly attractive one $a_\mathrm{KRb} = (-82.4\pm1.6) a_0$ (bottom row). Both $^{41}$K and $^{87}$Rb atoms are imaged as described in Fig. \ref{fig:TOF}.}
\end{figure}

\begin{figure}[t]
\includegraphics[width = 0.42\textwidth]{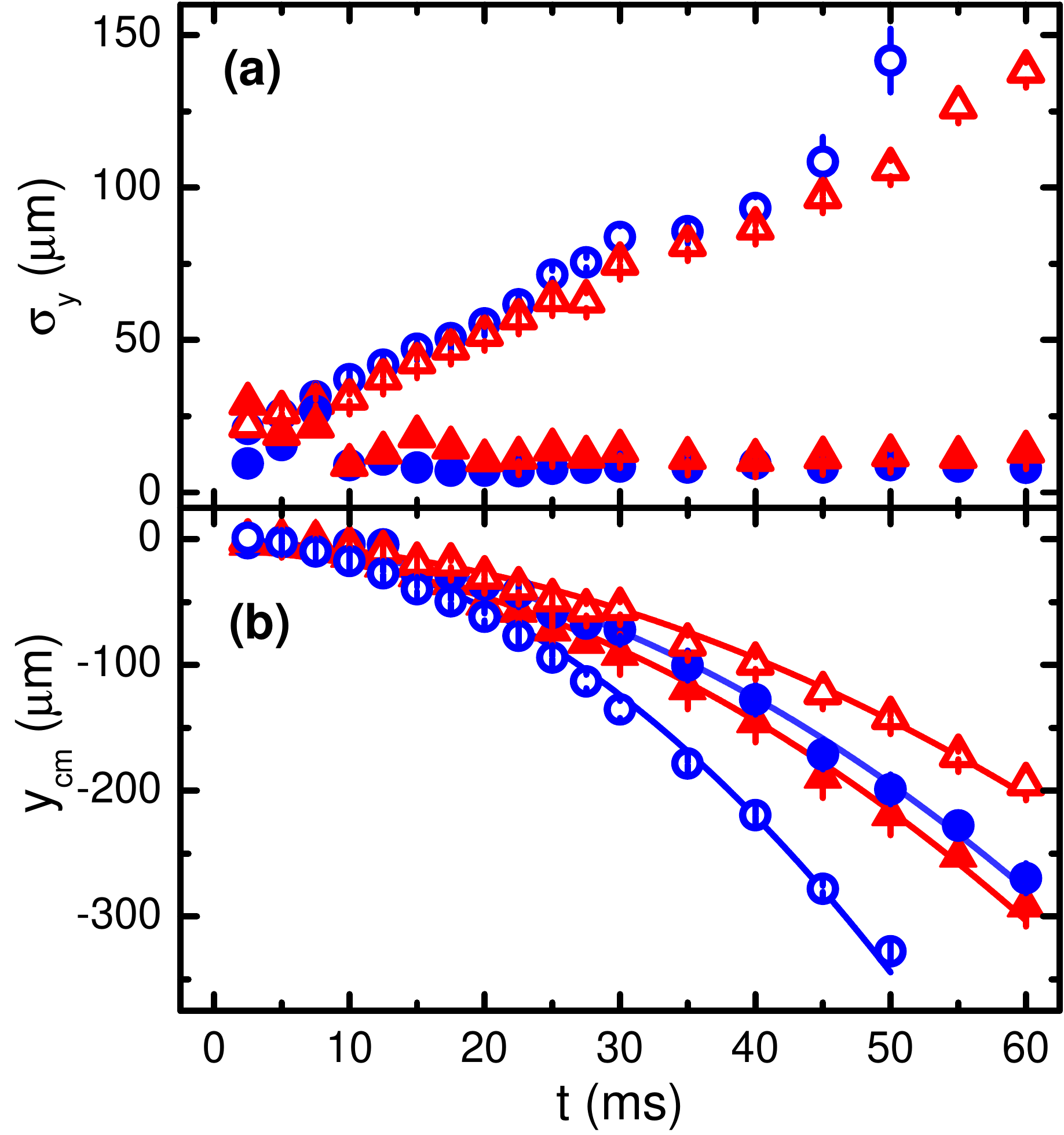}
\hspace{5mm}
\caption{\label{fig:GraficiEspansioneGuida} Evolution of the longitudinal size $\sigma_y$ (a) and the center-of-mass position $y_\mathrm{cm}$ along the waveguide (b) of $^{41}$K (blue open/closed circles) and $^{87}$Rb (red open/closed triangles) clouds, for $a_\mathrm{KRb} = (-11.4\pm1.4)/(-82.5\pm1.6)$~$a_0$.
Solid lines in (b) are fits with constant acceleration. The measured accelerations of the expanding clouds are $\alpha_{\mathrm{K}}=(0.276\pm0.025)$~m$/$s$^2$ and $\alpha_{\mathrm{Rb}}=(0.104\pm0.010)$~m$/$s$^2$ for $a_\mathrm{KRb} = (-11.4\pm1.4) a_0$. While for the bound components we find $\alpha_d=(0.152\pm0.017)$~m$/$s$^2$ ($\alpha_d=(0.151\pm0.022)$~m$/$s$^2$) from images of $^{41}$K ($^{87}$Rb).  The error bars are the standard deviation of, typically, five independent measurements.}
\end{figure}

To minimize the effect of gravity, we hold the atoms in a horizontal waveguide and let them evolve under the dominant effect of a magnetic field gradient along the $\hat{y}$ direction.
In Fig. \ref{fig:ImmaginiEspansioneGuida} we show {\it in-situ} images of the two condensates as a function of the evolution time $t$ in the waveguide. In this configuration we are able to detect the bound states for a longer time, since the center-of-mass displacement is less than $300\;\mu$m in $50$~ms.
For $\delta g>0$ (top), both species expand as $t$ increases, while for $\delta g<0$ (bottom) we observe the formation of self-bound states. Also here, as well as in free space, a fraction of the atoms remains unbound and follows the same dynamics of the weakly-attractive gas.
In Fig. \ref{fig:GraficiEspansioneGuida} we plot the size $\sigma_y$ and the center-of-mass position $y_\mathrm{cm}$ along the longitudinal direction. For $\delta g<0$ we fit the atomic density distribution with two Gaussians to distinguish the bound and unbound components.
For clarity, $\sigma_y$ and $y_\mathrm{cm}$ are shown only for the bound components, since the unbound fraction behaviour is very similar to the one observed for $\delta g>0$ (see Appendix \ref{AppendixC}).
The measured size for $\delta g<0$ (closed symbols) is constant and is comparable with the imaging resolution in the waveguide ($10\;\mu$m) \footnote{We use two different imaging systems: for measurements in free space (in the waveguide) the imaging beam propagates along the waveguide (crossed beam) and the optical resolution is $5\;\mu$m ($10\;\mu$m).}, while for $\delta g >0$ both species expand (open symbols). 
In the waveguide, the center-of-mass moves under the combined effect of a magnetic force and a residual component of the gravity, since the guide is tilted of $\theta\simeq 0.1-0.2^{\circ}$ with respect to the horizontal plane.
Assuming a constant acceleration (see Appendix \ref{AppendixC}), we extract the ratio of atom number of the two species in the bound state as
\begin{equation}
\label{eq:ratio}
\frac{N^d_\mathrm{K}}{N^d_\mathrm{Rb}}=\frac{m_\mathrm{Rb}(\alpha_d-\alpha_\mathrm{Rb})}{m_\mathrm{K}(\alpha_\mathrm{K}-\alpha_d)},
\end{equation}
where $\alpha_\mathrm{K}$ and $\alpha_\mathrm{Rb}$ are the accelerations of the unbound $^{41}$K and $^{87}$Rb clouds, and $\alpha_d$ is the acceleration of the bound state.
From $\alpha_{\mathrm{K}}$ and $\alpha_{\mathrm{Rb}}$ measured at $\delta g >0$ we find $N^d_\mathrm{K}/N^d_\mathrm{Rb} = 0.8(3)$, which is consistent with the predicted value of $\sqrt{g_\mathrm{Rb}/g_\mathrm{K}}=0.85$ \cite{Petrov}. We note that the center-of-mass motion allows a measurement of the atom number ratio which is not affected by the usual systematic uncertainty of $N$ measurements.
\subsection{Comparison with theory}
\label{subsec:Comparison}
\begin{figure}[t]
\subfigure{\includegraphics[width = 0.42 \textwidth]{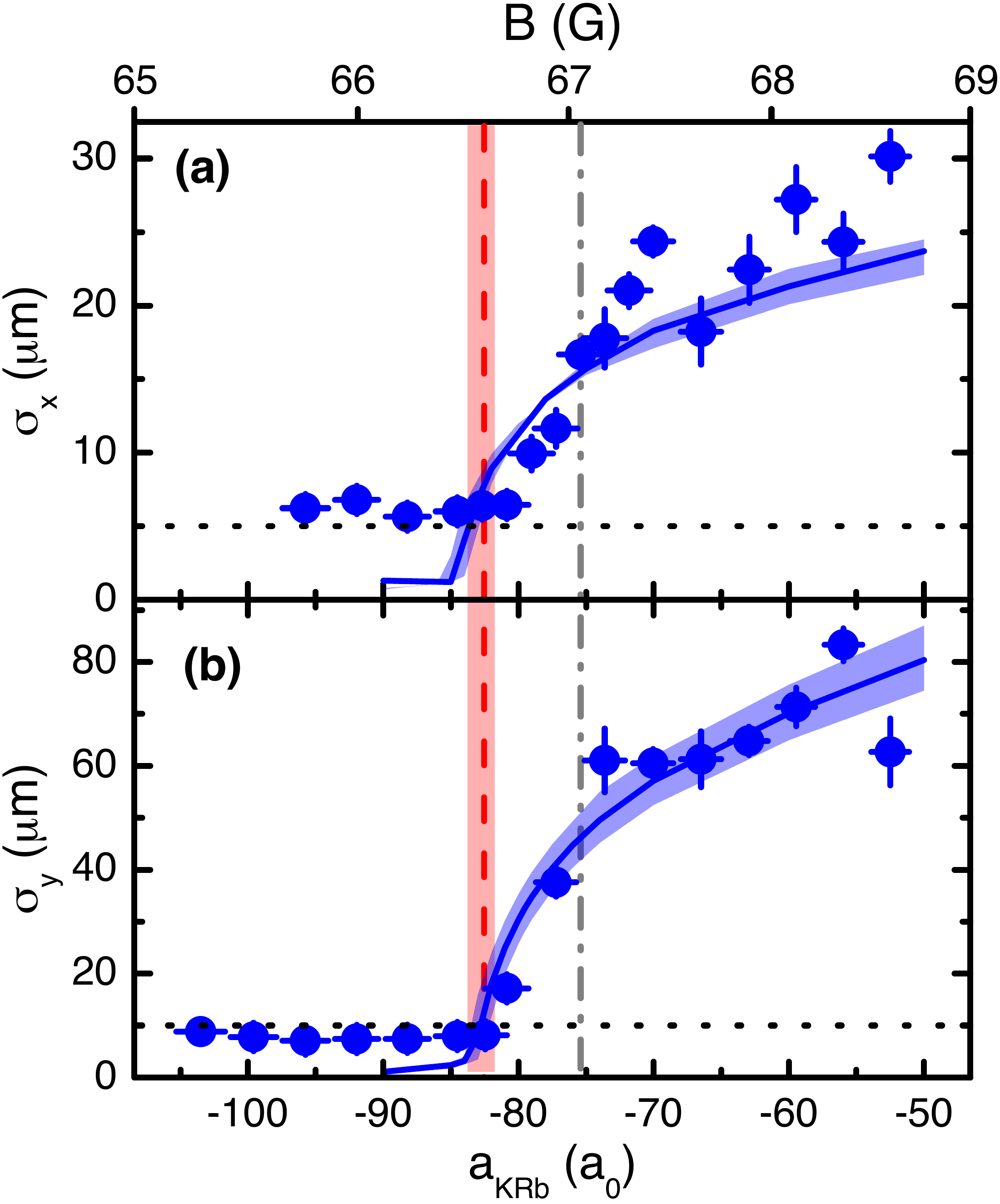}}
\hspace{5mm} 
\caption{\label{fig:Sigma_vs_a} (a) Size $\sigma_x$ of the $^{41}$K cloud expanding in free space as a function of the interspecies scattering length $a_\mathrm{KRb}$ after TOF $=26.5$~ms from the crossed trap. (b) Axial size $\sigma_y$ of $^{41}$K cloud as a function of the interspecies scattering length $a_\mathrm{KRb}$ after expansion in the waveguide of $t=45$~ms. The dashed red vertical line corresponds to the critical $a_\mathrm{KRb}$ for the existence of a droplet in the homogeneous case with $N_\mathrm{K}+N_\mathrm{Rb}=33000$, while the dash-dot gray vertical line corresponds to $\delta g = 0$. The solid lines are the numerical predictions for $N_\mathrm{K}+N_\mathrm{Rb}=33000$, where we fix $N_\mathrm{Rb}=\sqrt{g_\mathrm{K}/g_\mathrm{Rb}}\times N_\mathrm{K}$. The colored areas indicate the systematic relative uncertainty on the experimental atom number of 30 \%. We do not appreciate a substantial change of the average total atom number in the explored range of $a_\mathrm{KRb}$. The horizontal dotted lines show the imaging resolution of $5\;\mu$m and $10\;\mu$m for the measurements in free space and in the waveguide, respectively. Vertical error bars: standard deviation of, typically, five independent measurements.}
\end{figure}

Both in free space and in the waveguide, we study the full range of attractive interactions.
In Fig.~\ref{fig:Sigma_vs_a} we plot the width $\sigma_x$ of $^{41}$K in free space after TOF $=26.5$~ms (Fig.~\ref{fig:Sigma_vs_a}a) and $\sigma_y$ after $t = 45$~ms of expansion in the waveguide (Fig.~\ref{fig:Sigma_vs_a}b). In both cases we observe a transition from an expanding gas for $a_\mathrm{KRb} \gtrsim - 82 a_0$ to a localized self-bound state for strong attractive interactions ($a_\mathrm{KRb} \lesssim - 82 a_0$), whose width is below the optical imaging resolution. In Fig.~\ref{fig:Sigma_vs_a}, we also show the MF threshold for collapse (dash-dot gray line) and the critical scattering length to enter the droplet regime with our experimental atom number (dashed red line) \cite{Petrov}. Both lines are calculated for a homogeneous system. In the presence of the radial confinement a stable bound state may exist also in-between the two vertical lines. This is due to the combined effect of the LHY term and the dispersion along the guide giving rise to the formation of ``solitonic'' solutions \cite{Shabat72,Cappellaro2018}.

In order to compare the experimental results with theoretical predictions \cite{Petrov,Ancilotto2018}, we have performed numerical simulations by using two coupled generalized time-dependent GP equations including the LHY term, taking into account the actual experimental preparation (see Appendix \ref{AppendixB} for details). In Fig. \ref{fig:Sigma_vs_a} the colored areas correspond to the numerical predictions including the experimental uncertainty on the atom number. Considering the finite imaging resolution, we find that the theory well reproduces the observed behavior. 
From the simulations in free space (Fig.~\ref{fig:Sigma_vs_a}a) we find that, in the strongly attractive regime, almost spherical droplets form \cite{Petrov}. As expected, the critical scattering length to enter the droplet regime depends on the atom number. We find that this threshold  actually occurs at larger values of $|a_\mathrm{KRb}|$ with respect to the homogeneous case (see dashed red line in Fig.~\ref{fig:Sigma_vs_a}). This is due to the additional kinetic energy cost associated to the droplet surface, whose effects are more important for small droplets, as those formed in the experiment.

As for the waveguide dynamics, we observe expanding clouds in-between the two vertical lines instead of the predicted ``solitonic'' bound states. We attribute this to a twofold effect: \textit{(i)} we adjust the initial size of the mixture to match the droplet size (smaller than the one of the solitonic ground state); \textit{(ii)} the dynamics is triggered by a non adiabatic process. However, droplets formed for very large attractive scattering lengths ($|a_\mathrm{KRb}| \gtrsim 95 a_0$), as already observed in previous experiments \cite{Cheiney2018}, eventually decay into ``solitonic'' solutions due to the decrease of the atom number induced by three body losses. In the waveguide, indeed, atomic losses do not affect the existence of self-bound states allowing their observation for a wider range of $a_\mathrm{KRb}$. In free space, instead, as confirmed by numerical simulations with three body losses (Appendix \ref{AppendixB}), we are limited to $|a_\mathrm{KRb}|\lesssim 95 a_0$. This corresponds to a broader range of $\delta g$ with respect to the case of $^{39}$K droplets, pushing the investigation far away from the MF collapse.
\section{Conclusions}
In conclusion, we have observed quantum droplets in a mixture of two atomic species experiencing different potentials, providing a clear picture of their formation and evolution both in free space and under radial confinement. This proves that the presence of identical traps for the two components is not essential to the droplets production. We have found  evidence that the lifetime of our droplets is substantially longer than those in $^{39}$K, as expected. The droplet density obtained by our numerical simulations results in a lifetime up to $ 400$~ms. 
This opens new perspectives for future studies of the droplets peculiar properties \cite{Petrov,Ancilotto2018} and of their collective modes \cite{Cappellaro2018}. Indeed, the lower density of our sample allows for a larger ratio $\tau_\mathrm{life}/\tau \sim n^{-1}$, with $\tau_\mathrm{life}$ and $\tau$ being respectively the lifetime and the characteristic time scale of the droplet dynamics (Appendix \ref{AppendixB}). Other interesting directions could be the creation of droplets in reduced dimensionality \cite{Petrov2016,Santos2017,Astrakharchik2018,Zin2018,Parisi2019}, where quantum fluctuations become easily dominant, and the investigation of beyond LHY corrections \cite{Cikojevic2019} and other stabilization mechanisms \cite{Adhikari2017,Gautam_2019}. Finally, the nucleation of vortex-antivortex ring pairs \cite{Kartashov2018,Ancilotto_He,Escartin2019} or more exotic scenarios, like droplets clusters \cite{Kartashov2019}, could be studied with rotating droplets.

\begin{acknowledgments} 
We acknowledge inspiring discussions with M. Fattori. We gratefully thank M. Inguscio for support and A. Simoni for collisional calculations.
This work was supported by the European Commission through FET Flagship on Quantum Technologies - Qombs Project (Grant No. 820419) and by Fondazione Cassa di Risparmio Firenze through project "SUPERACI-Superfluid Atomic Circuits".
M. M. acknowledges support by the Spanish Ministry of Science, Innovation and Universities and the European Regional Development Fund FEDER through Grant No. PGC2018-101355-B-I00 (MCIU/AEI/FEDER,UE), and the Basque Government through Grant No. IT986-16.
\end{acknowledgments} 

\appendix

\section{FESHBACH RESONANCES, MAGNETIC MOMENTS AND THREE BODY LOSSES}
\label{AppendixA}
\begin{figure}[h]
\includegraphics[width = 0.4 \textwidth]{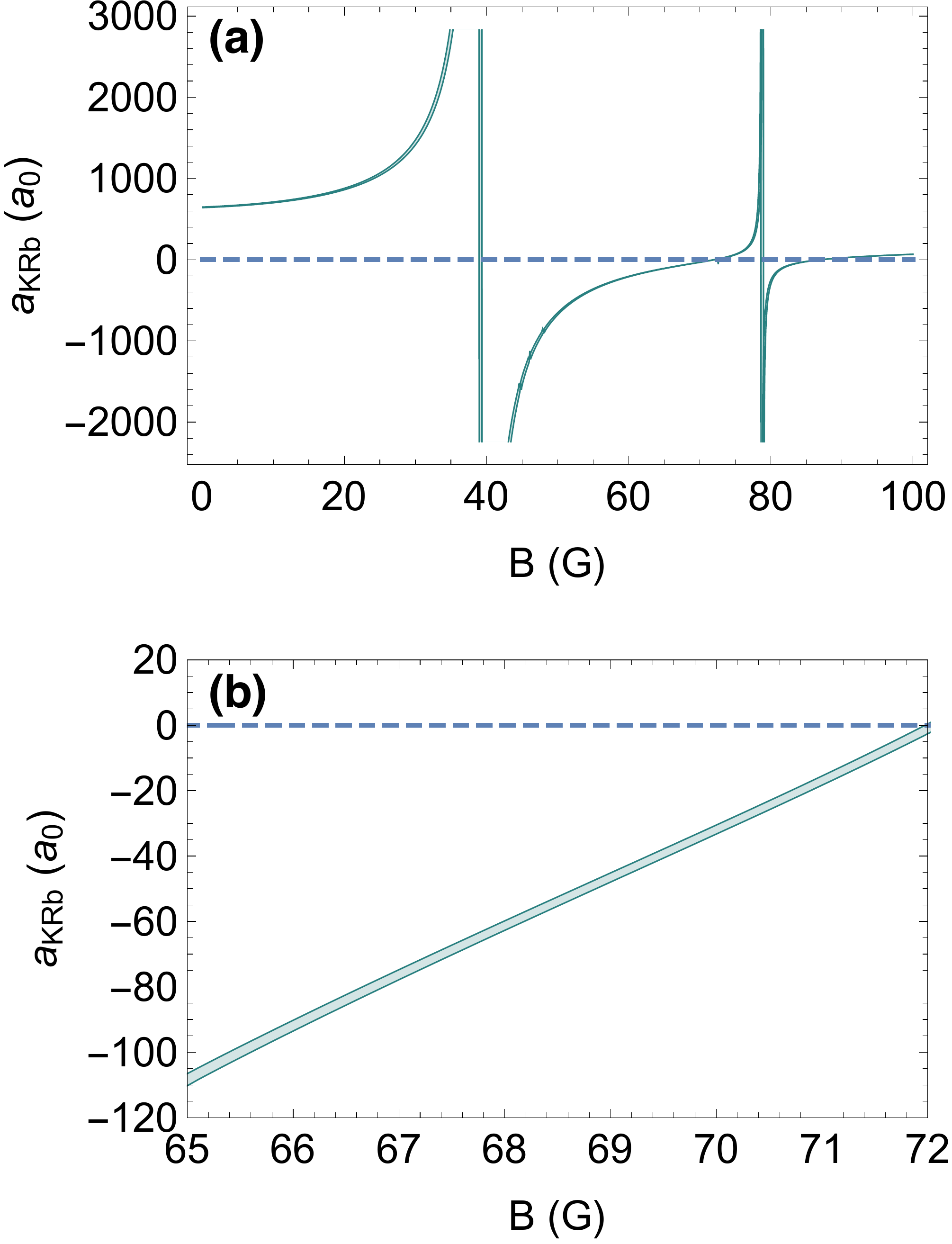}
\caption{\label{fig:Fesh} (a) Calculated interspecies scattering length $a_{\mathrm{KRb}}$ for $^{41}$K and $^{87}$Rb in the $\left|F=1,m_{F}=1\right\rangle$ state as a function of the magnetic field $B$. Both the prediction of Ref. \cite{Simoni2008} and Ref. \cite{Thalhammer2009} are shown by the thickness of the curve. (b) Zoom in the range of magnetic field used in the experiment.}
\end{figure}

In Fig. \ref{fig:Fesh}, we show the behaviour of the interspecies scattering lengths $a_{\mathrm{KRb}}$  for $^{41}$K and $^{87}$Rb in the $\left|F=1,m_{F}=1\right\rangle$ state as a function of the magnetic field $B$. Two Feshbach resonances around 40~G and 80~G can be seen. The values of $a_{\mathrm{KRb}}$ predicted by the two collisional models in Ref.\cite{Simoni2008} and in Ref.\cite{Thalhammer2009} slightly differ by few $a_0$, as shown by the thickness of the curve in Fig.~\ref{fig:Fesh} (b).
All the values of $a_\mathrm{KRb}$ given in this work are the average of the two models, with uncertainty equal to the half-deviation that dominates over the one due the magnetic field calibration.

\begin{figure}[h]
\includegraphics[width = 0.45 \textwidth]{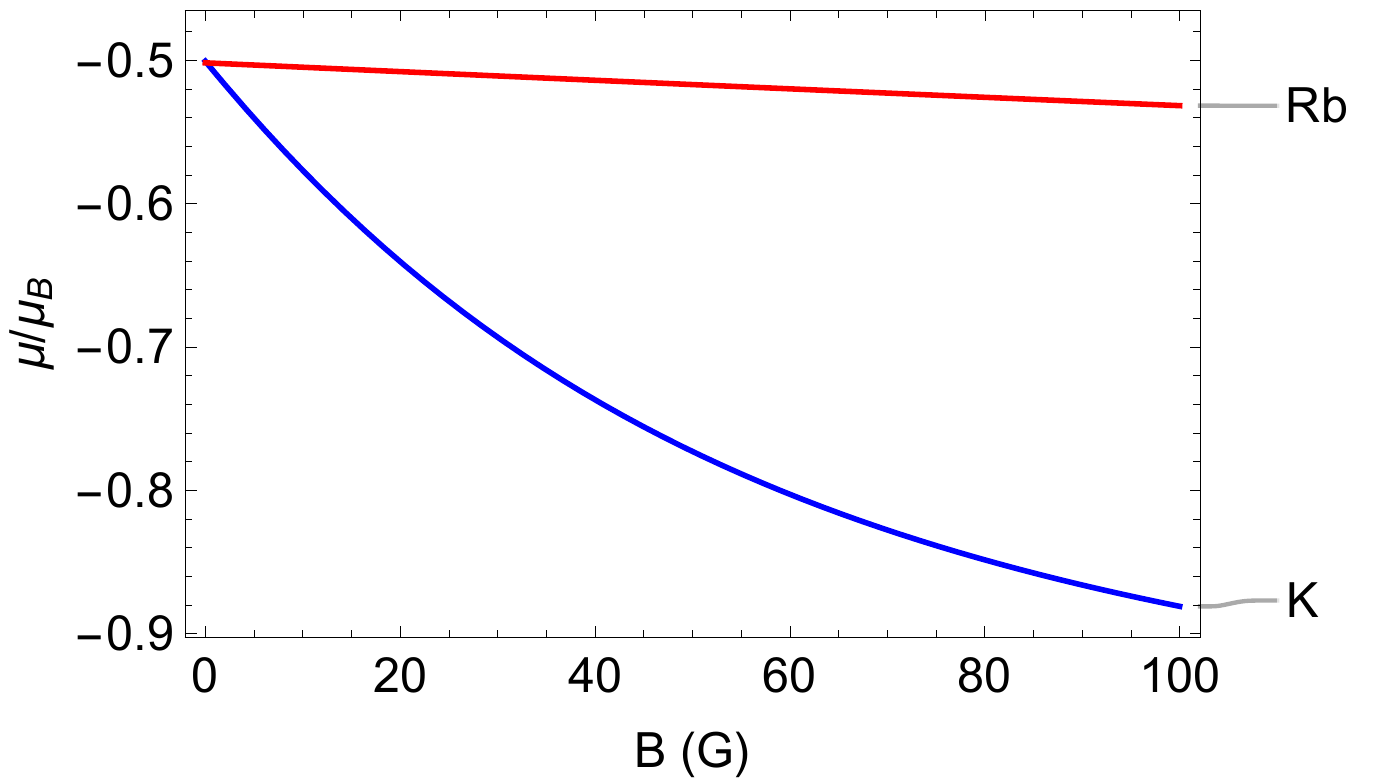}
\caption{\label{fig:mu} Calculated magnetic moment $\mu$ (in unit of the Bohr magneton $\mu_\mathrm{B}$) of the $\left|F=1,m_{F}=1\right\rangle$ state for both $^{41}$K and $^{87}$Rb as a function of the magnetic field $B$. 
}
\end{figure}

In the experiment we produce the double condensate at $a_{\mathrm{KRb}}\simeq 0$ and then we tune the magnetic field in the range $65~$G$\lesssim B \lesssim 69~$G to access the attractive inter-species interaction regime ($-100~a_0\lesssim a_{\mathrm{KRb}}\lesssim -50~a_0$). In particular, we use two linear ramps: from $71$~G to $68.5$~G in $20$~ms and then to the final value in other $10$~ms.
At these magnetic fields, the Zeeman effect is no longer linear (especially for $^{41}$K) and the magnetic moments $\mu_{\mathrm{K}}$ and $\mu_{\mathrm{Rb}}$ of the hyperfine state of  $^{87}$Rb and $^{41}$K in our mixture vary as shown in Fig. \ref{fig:mu}. In particular, in the range of $B$ used in the experiment the two magnetic moments are quite different, enabling a different magnetic control of the two atomic species, while their relative variations
$\Delta \mu_{\mathrm{K}}/\mu_{\mathrm{K}}$ and $\Delta \mu_{\mathrm{Rb}}/\mu_{\mathrm{Rb}}$ are below 4\% and therefore neglected.
The difference between the two magnetic moments is exploited both to superimpose the two condensates by means of a magnetic gradient $b_z$, and to induce a different acceleration of the unbound and bound components.


As mentioned in Section \ref{subsec:Comparison}, we include in our analysis a loss term due to inelastic three-body collisions. In general, the three-body losses coefficient $K_3$ depends on the scattering length, scaling as $a^4$ in vicinity of a Feshbach resonance. In our case, we explored a very limited range of magnetic fields and scattering lengths $a_\mathrm{KRb}$, therefore we measured the atoms lifetime to obtain the rate of inelastic three-body collisions at $a_{\mathrm{KRb}}=(-77.8\pm1.6) a_0$. We measured the lifetime of the $^{41}$K sample, since {\it (i)} the number of the minority species is more sensitive to losses; {\it (ii)} the dominant three-body losses are due to K-Rb-Rb collisions \cite{wacker2016}, yielding a purely exponential decay for $^{41}$K atoms. We measured a $1/e$-lifetime $\tau_\mathrm{K}=122(11)$~ms with an estimated $^{87}$Rb peak density $n_{0,\mathrm{Rb}}=(6 \pm 2)\cdot 10^{20}$~m$^{-3}$. From $1/\tau_\mathrm{K} = (K_3/2!) n^2_{0,\mathrm{Rb}} \xi$, with 
$\xi = \int n_{\mathrm{Rb}}^2 n_{{\mathrm K}} dV/(n^2_{0,\mathrm{Rb}}N_{\mathrm{K}})$, we obtain $K_3 = (7\pm 4)\cdot 10^{-41}$~m$^6$/s with an uncertainty due to that of $n_{0,\mathrm{Rb}}$. This must be compared to published values for larger scattering lengths: $K_3 \simeq 10^{-38}$~m$^6$/s at $a_\mathrm{KRb}\simeq -120 a_0$ \cite{wacker2016} and $K_3 \simeq 3\cdot 10^{-39}$~m$^6$/s at $a_{\mathrm{KRb}}\simeq -400 a_0$ \cite{kato2017}.

\section{THEORETICAL ANALYSIS}
\label{AppendixB}
The Gross-Pitaevskii (GP) energy functional for a Bose mixture, including both the mean field term and the Lee-Huang-Yang (LHY) correction accounting for quantum fluctuations in the local density approximation, reads \cite{Petrov,Ancilotto2018}

%
%
\begin{widetext}
\begin{equation}
E = \sum_{i=1}^{2}\int d\bm{r}\left[{\hbar^2 \over 2m_i}|\nabla \psi_{i}(\bm{r})|^2  + V_{i}(\bm{r}) n_{i}(\bm{r})\right]
+\frac{1}{2}\sum_{i,j=1}^{2}g_{ij}\int d\bm{r}n_{i}(\bm{r})n_{j}(\bm{r})
+\int d\bm{r}{\cal E} _{\rm LHY}(n_1(\bm{r}),n_2(\bm{r}))\,,
\label{eq:energyfun}
\end{equation}
\end{widetext}
where $V_i(\bm{r})$ and $n_i(\bm{r})=|\psi _i(\bm{r})|^2$ represent the external potential and the density of each component ($i=1$ for $^{41}$K, $i=2$ for $^{87}$Rb) and the LHY correction is \cite{Petrov}
\begin{align}
{\cal E} _{\rm LHY} &= {8\over 15 \pi^2} \left(\frac{m_1}{\hbar^2}\right)^{3/2}
\!\!\!\!\!\!(g_{11}n_1)^{5/2} 
f\left(\frac{m_2}{m_1},\frac{g_{12}^2}{g_{11}g_{22}},\frac{g_{22}\,n_2}{g_{11}\,n_1}\right)
\nonumber
\\
&\equiv \kappa (g_{11}n_1)^{5/2}f(z,u,x).
\end{align}
Here $f(z,u,x)>0$ is a dimensionless function, whose expression for $z\neq 1$ and $u=1$ can be found in \cite{Ancilotto2018}. Following \cite{Petrov, Ancilotto2018}, we also consider this function at the mean-field collapse $u=1$, $f(z,1,x)$.
We note that the actual expression for $f$ can be fitted very accurately with the same functional form of the homonuclear case ($m_{1}=m_{2}$) \cite{Minardi}
\begin{equation}
f\left(87/41,1,x\right)\simeq \left(1+\alpha x\right)^{\beta}
\end{equation}
where $\alpha$ and $\beta$ are fitting parameters that in general depend on the value of the mass ratio $m_1/m_2$. 
For the present case we find $\alpha\simeq 1.554$ and $\beta\simeq 2.506$, which are very close to the values of the approximated formula proposed in Ref. \cite{Minardi}, which shows that only $\alpha$ is a function of the mass ratio, $\alpha=(m_2/m_1)^{3/5}=1.571$, whereas the value of $\beta$ is universal, $\beta=5/2$.

Minimization of the action associated to Eq.~(\ref{eq:energyfun}) leads to the following \textit{generalized} GP equations (Euler-Lagrange equations)

\begin{equation}
i \hbar {\partial \psi _i \over \partial t} =
\left[-{\hbar^2 \over 2m_i}\nabla ^2 + V_i + \mu_{i}(n_1,n_2) \right]\psi _i
 \, , 
\label{eq:gpe}
\end{equation}
where  
\begin{equation}
\mu_{i}\equiv \frac{\delta E}{\delta n_{i}} = g_{ii}n_i+g_{ij}n_j+ \frac{\partial {\cal E}_{\rm LHY}}{\partial n_i}\quad \,( j \ne i) \,,
\label{eq:chempot}
\end{equation}
and
\begin{align}
\frac{\partial {\cal E} _{\rm LHY}}{\partial n_1} &=
\kappa g_{11}(g_{11}n_{1})^{3/2}\left({5\over 2}f-x{\partial f\over \partial x}\right)\,,
\\
\frac{\partial {\cal E} _{\rm LHY}}{\partial n_2} &=
\kappa g_{22}(g_{11}n_{1})^{3/2}{\partial f\over \partial x}\,.
\end{align}
The above equations are solved by mapping the system (densities, wave functions, differential operators, etc.) on discrete equally spaced cartesian grids. The differential operators are represented by a 13-point discretization. The stationary GP equation is solved by imaginary time evolution. Dynamical equations have been solved by using Hamming's predictor-modifier-corrector method, initiated by a fourth-order Runge-Kutta-Gill algorithm \cite{Ralston1960}.
The effect of three-body losses on the real-time dynamics of the mixture is simulated by adding to the energy functional in Eq.~(\ref{eq:energyfun}) a dissipative term
$-(i/2)\hbar K_3 \int d\bm{r} n_1(\bm{r},t)n_2(\bm{r},t)^2$
(with $K_3=7\cdot 10^{-41}$ m$^6$/s, see Appendix \ref{AppendixA})
which accounts for the dominant recombination channel, \textit{i.e.} K-Rb-Rb.

We performed two different series of simulations of the experiment for the TOF in free space and for the expansion in the waveguide. In the first case, the ground state of the mixture is produced in two concentric harmonic potentials (for the two condensates) approximating the actual crossed dipole trap. Then the potential is switched off abruptly and an evolution in free space is followed taking into account three-body losses.
For the simulation in the waveguide, we adopt a complete description of the potential experienced by the two species including the optical, magnetic and gravitational potentials. This allows us to reproduce the effect of a residual magnetic field and a differential gravitational sag of the two components.
The results reported in Fig. 4 (b) are obtained neglecting three-body losses after checking that their contribution is negligible in the range of parameters investigated.
All results of the simulations shown in Fig. 4 have been performed with $N_{\mathrm{Rb}}/N_{\mathrm{K}}=\sqrt{g_{\mathrm{K}}/g_{\mathrm{Rb}}}$. Any unbalance from this ratio leads to an unbound expanding fraction which does not affect the bound fraction, but heavily affects computation time requiring a larger spatial grid. The simulations show the formation of self-bound states even when the trapping potentials of the two condensates are not identical and completely overlapped. A detailed investigation of this effect goes beyond the scope of the present work.

The equilibrium density of a droplet can be obtained by the condition $P=-\epsilon+\sum_{i}({\partial\epsilon}/{\partial n_i})\,n_i=0$, where $P$ is the pressure and $\epsilon$ is the energy density. For the heteronuclear case this condition gives:
%
%
\begin{align}
n^{0}_{1}&=\frac{25\pi}{1024\, a_{11}^3}\frac{\delta a^2}{a_{11}a_{22}}
\frac{\left(m_1+m_2\right)^{2}}{4m_1m_2}
\left(
1+\sqrt[10]{\frac{m_2}{m_1}}\sqrt{ \frac{a_{22}}{a_{11}}} 
\right)^{-5}
\nonumber\\
n^{0}_{2}&=n^{0}_{1}\sqrt{\frac{a_{11}m_2}{a_{22}m_1}}.
\label{eq:density1}
\end{align}
Here $a_{11}$, $a_{22}$ and $a_{12}$ are the intra and interspecies scattering lengths and $\delta a\equiv a_{12}-a^{c}_{12}$, with $a^{c}_{12}$ the mean-field threshold for collapse.
From Eq.~(\ref{eq:density1}) we estimate that, in our mixture, the homogeneous equilibrium density $n^{0}_{1}$ varies from $\sim2\cdot 10^{14}$ atoms/cm$^3$ to $\sim2\cdot 10^{15}$ atoms/cm$^3$ for $-6~a_0<\delta a< -20~a_0$, corresponding to the range for which we observe a droplet state (Fig. 4 (a)). Furthermore, for $^{41}$K and $^{87}$Rb, the terms containing the masses are $\sim$ 1 and Eq.~(\ref{eq:density1}) is not substantially different from the equation for the homonuclear case given in \cite{Petrov}:
\begin{equation}
n^{0}_{1}=\frac{25\pi}{1024\, a_{11}^3}\frac{\delta a^{2}}{a_{11}a_{22}}
\left(
1+\sqrt{ \frac{a_{22}}{a_{11}} } 
\right)^{-5}
\label{eq:homo}
\end{equation}
From Eq.~(\ref{eq:homo}), it is evident that upon increasing all scattering lengths by a factor $\gamma>1$, the densities decrease as $\gamma^{-3}$.
To compare the equilibrium density of droplets formed either by $^{41}$K-$^{87}$Rb mixture or with two spin states $^{39}$K, we use Eq.~(\ref{eq:density1}) and Eq.~(\ref{eq:homo}), respectively.
In the first case $a_{11}$ and $a_{22}$ are larger than those of $^{39}$K (for which $a_{22}$ is tuned through a Feshbach resonance, while $a_{11}$ and $a_{12}$ are constant), resulting in a droplet density lower by approximately an order of magnitude at $\delta a$ of a few $-a_0$. As already mentioned in the main text, this also results in a longer lifetime ($\tau_\mathrm{life} \sim n^{-2}$) and a larger ratio $\tau_\mathrm{life}/\tau \sim n^{-1}$, with $\tau$ being the characteristic time scale of the droplet \cite{Petrov}. 

\section{WAVEGUIDE DYNAMICS}
\label{AppendixC}
\subsection{Equations of motion}
In this section we derive the equations of motion for both atomic species confined in the waveguide. 
We start by considering the case in which they are unbound.
The \textit{anti-confining} force experienced by the $i$-th atomic species along the $\hat{y}$ axis is mainly determined by the magnetic potential $U_{y,i}$. This is given by the Feshbach field $B_F$ and the overlapping magnetic field gradient, which are produced by two different pairs of Helmholtz and anti-Helmholtz coils, along the $\hat{z}$ axis, respectively. Thus, we have
\begin{equation}
U_{y,i}=-\left|\mu_i\right|\sqrt{B_F^2+\left(b_y y\right)^2},
\label{magnetic_potential}
\end{equation}
where $\mu_i$ is the magnetic moment and $b_y$ is the $\hat{y}$ component of the magnetic field gradient.
For $|b_y y|\ll B_F$, 
which is fulfilled in our experimental conditions, $U_{y,i}$ can be approximated as  
\begin{equation}
U_{y,i}\simeq-\left|\mu_i\right|B_F\left[1+\frac{1}{2}\left(\frac{b_y y}{B_F}\right)^{2}\right],
\label{magnetic_potential_approximation}
\end{equation}
where we neglect the terms in $\left(b_y y\right)/ B_F$ higher than the second order.
If we define $\omega_i^2=\left|\mu_i\right|/(m_i B_F)b_y^2 $, with $m_i$  the atomic mass, the equation of motion is reduced to
\begin{equation}
\ddot{y_i}-\omega_i^{2}y_i=0,
\label{eq:motion}
\end{equation}
where ${y_i}$ is the center-of-mass position of the atomic cloud and $\ddot{y_i}$ 
its acceleration. The solution of equation Eq.~(\ref{eq:motion}) is 
\begin{equation}
y_i\left(t\right)=A_1\,\mathrm{cosh}\left(\omega_i t\right)+A_2\, \mathrm{sinh}\left(\omega_i t\right),
\label{eq:solution}
\end{equation}
where the constant factors $A_1$ and $A_2$ are determined by the initial conditions. For short times ($t\ll \min\{\omega_{i}^{-1}\})$, Eq.~(\ref{eq:solution}) can be approximated by its Taylor series expansion as
\begin{equation}
y_i\left(t\right) \simeq y_i\left(0\right)\left[1+\frac{1}{2} \left(\omega_i t\right)^{2}\right]+\dot{y_i}\left(0\right) t
\label{eq:solution_approx}
\end{equation}
which corresponds to a uniformly accelerated motion, with constant acceleration $\alpha_{i}\equiv y_i\left(0\right)\omega_{i}^{2}$.
Eq.~(\ref{eq:solution_approx}) is a good approximation of the observed center-of-mass dynamics, and, indeed, it has been used for fitting the experimental data discussed in the main text.
Here, $y_i(0)$ is the initial distance of the atomic cloud from the maximum of the magnetic potential and $\dot{y_i}\left(0\right)$ is the initial velocity.
The value of $y_i(0)$ is not zero since the position of the atoms in the dipole trap is displaced from the symmetry axis of the magnetic field coils. We found that setting $\dot{y_i}\left(0\right)=0$ does not substantially affect the fitting results. Further we verified that, with $\dot{y_i}=0$, for evolution times up to 60~ms the error in $y_i$ made using Eq.~(\ref{eq:solution_approx}) instead of Eq.~(\ref{eq:solution}) is less than 1 \%. 

We now treat the case of the bound state
assuming that it is formed by $N^{d}_{i}$ atoms in each component ($i=1,2$), then its equation of motion is given by
\begin{equation}
\ddot{y}_{d} \sum_{i=1}^{2} N^{d}_{i} m_{i}=\sum_{i=1}^{2} F_{i},
\label{eq:motion_bound}
\end{equation}
where $y_{d}$ is the center-of-mass of the droplet, and $F_{i}$ is the total (constant) force acting on each component. This force can also be expressed in terms of the center-of-mass acceleration $\alpha_{i}$ that the $i-$th component would have in the absence of the other species, namely $F_{i}=N^{d}_{i}m_{i}\alpha_{i}$. Then, Eq.~(\ref{eq:motion_bound}) can be rewritten as
\begin{equation}
\ddot{y}_{d} \sum_{i} N^{d}_{i} m_{i}=\sum_{i} N^{d}_{i}m_{i}\alpha_{i},
\end{equation}
from which it is straightforward to get Eq.~(2). 

\subsection{Experimental fit of the unbound state dynamics}
\begin{figure}[t]
\includegraphics[width = 0.45 \textwidth]{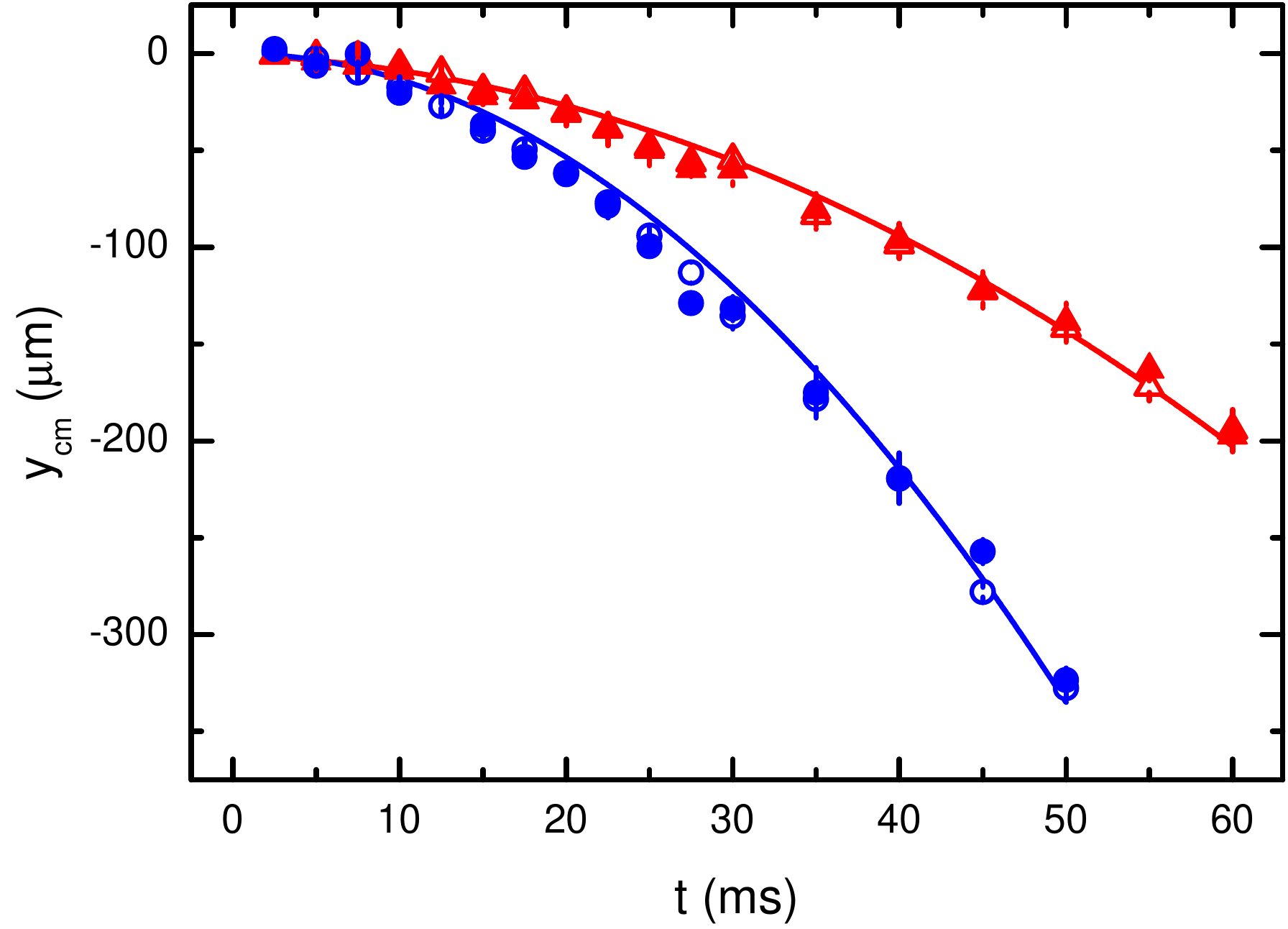}
\hspace{5mm}
\caption{\label{fig:CMEspansioneGuidaSI} Evolution of the center-of-mass position $y_\mathrm{cm}$ along the waveguide for $a_\mathrm{KRb} = (-11.4 \pm 1.4) a_0$ of $^{41}$K (blue open circles) and $^{87}$Rb (red open triangles), and for the expanding fractions at $a_\mathrm{KRb} = (-82.5 \pm 1.6) a_0$ of  $^{41}$K (blue closed circles) and $^{87}$Rb (red closed triangles). Solid lines are fit with constant acceleration for $a_\mathrm{KRb} = (-82.5 \pm 1.6) a_0$. The fit results are: $\alpha_{\mathrm{K}}=(-0.27 \pm 0.07)$ m$/$s$^2$ and $\alpha_{\mathrm{Rb}}=(-0.104 \pm 0.013)$ m$/$s$^2$.}
\end{figure}

In the droplet regime, we generally observe a portion of atoms left over the bound state. We ascribe this effect to the presence of a thermal fraction in both atomic species and an excess of $^{87}$Rb atoms with respect to the droplet atom ratio. Thus, to distinguish the bound and unbound parts, we fit the atomic density distribution with a two component Gaussian function.
In Fig. 4 we have shown only the results obtained for the bound portion. Here we report also the evolution of the center-of-mass position $y_\mathrm{cm}$ of the unbound fraction for $\delta g <0$ finding the same behaviour of the weakly attractive mixture (Fig. \ref{fig:CMEspansioneGuidaSI}).

Following the same strategy described in the main text, from Eq.~(2), where now $\alpha_\mathrm{K}$ and $\alpha_\mathrm{Rb}$ are the accelerations of the unbound $^{41}$K and $^{87}$Rb components at $\delta g<0$, we obtain the atom number ratio of the two species in the bound state: $N^{d}_\mathrm{K}/N^{d}_\mathrm{Rb} = (0.8 \pm 0.7)$, again consistent with the expected value \cite{Petrov}.

\bibliography{Biblio}

\end{document}